# Novel dense crystalline packings of ellipsoids

Weiwei Jin[1], Yang Jiao[2], Lufeng Liu[1], Ye Yuan[1] & Shuixiang Li[1,*]

[1]Department of Mechanics and Engineering Science, Peking University, Beijing 100871, China

[2]Materials Science and Engineering, Arizona State University, Tempe, AZ 85287, USA

*E-mail: lsx@pku.edu.cn

**Abstract**

An ellipsoid, the simplest non-spherical shape, has been extensively used as models for elongated building blocks for a wide spectrum of molecular, colloidal and granular systems[1-4]. Yet the densest packing of congruent hard ellipsoids, which is intimately related to the high-density phase of many condensed matter systems, is still an open problem. We discover a novel dense crystalline packing of ellipsoids containing 24 particles with a quasi-square-triangular (SQ-TR) tiling arrangement, whose packing density $\phi$ exceeds that of the SM2 crystal[5] for aspect ratios α in (1.365, 1.5625), attaining a maximal $\phi \approx 0.75806\ldots$ at $α = 93/64$. We show that SQ-TR phase is thermodynamically stable at high densities over the aforementioned α range and report a novel phase diagram for self-dual ellipsoids. The discovery of SQ-TR crystal suggests novel organizing principles for non-spherical particles and self-assembly of colloidal systems.

Dense packings of hard particles have been widely used as models for a variety of condensed matter systems, including glasses, crystals, heterogeneous materials, and granular media[1-4]. The optimal (maximally dense) packing arrangement is also intimately related to the crystalline phase of the associated particle system and its high-density phase behavior[3]. The pursuit of the optimal packing for given particle shape has long been an intriguing and challenging problem in discrete geometry[6]. Indeed, it took almost four centuries to prove the famous Kepler's conjecture[6] posed in 1611 that the densest packing of spheres is the face-centered cubic (FCC) lattice with a packing density $\phi = \pi/\sqrt{18} \approx 0.7405$ [7]. For congruent nonspherical particles that do not tile three-dimensional (3D) Euclidean space $R^3$, the optimal packing problem is only solved for infinite cylinders[8] and rhombic dodecahedra with a clipped corner[9]. So far no rigorous mathematical theory has been put forth for constructing or proving optimal packings of nonspherical particles. Recently, the discovery of the candidate optimal packings is significantly advanced via computer simulations. Numerical methods including the molecular dynamics method[10,11], Monte Carlo methods[12-15] and the periodic divide and concur (PDC) algorithm[16] have been employed to suggest the candidates for the maximal-density packings in $R^3$ for a wide spectrum of nonspherical hard particles including ellipsoids[5], superballs[17,18], lens-shaped particles[19], and polyhedra[13-15,20-24], to name but a few. No counterexample of the Ulam's conjecture[25] that all convex shapes in $R^3$ pack better than spheres has been found yet.

An ellipsoid, the simplest non-spherical shape, can be obtained from a sphere through an affine transformation. The ratios of the semiaxes of an ellipsoid can be expressed as *a:b:c=α:α$^β$:1* (*a≥b≥c*), where *α* is the aspect ratio, and *β* (0≤*β*≤1) is the skewness. When *β* = 1/2, the ellipsoid is neither prolate (*β*=0) nor oblate (*β*=1), and is referred to as the self-dual ellipsoid. Dense packings of congruent ellipsoids have been extensively studied in the past decades, and ellipsoid system is shown to exhibit a much richer phase behavior than that of spheres. Donev *et al.*[5] found a simple monoclinic crystal with two ellipsoids of different orientations per unit cell (SM2), which is denser than the stretched FCC (sFCC) crystal for all aspect ratios expect for the sphere point. The density of the SM2 packing only depends on the aspect ratio $\alpha$, and is independent of the skewness $\beta$. For prolate ellipsoids of revolution, the SM2 phase is more stable than the sFCC structure for all densities above the solid-nematic coexistence point for $\alpha \geq 2.0$[26], which revised the Frenkel-Mulder phase diagram[27] of hard ellipsoids of revolution ($\beta=0, 1$). Further details on the phase diagram of ellipsoids of revolution were obtained by decompressing densest known SM2 crystals through replica exchange Monte Carlo simulations[28]. Furthermore, two additional liquid crystalline phases, namely the discotic phase and the biaxial phase, are observed for biaxial ellipsoids[29-31]. For the disordered branch, recent numerical[32-36] and experimental[36-39] studies have shown that disordered jammed packings of ellipsoids possess fewer contacts than that predicted by the isocounting conjecture[32], which states that to constrain a system two contacts per degree of freedom are required. These hypostatic packings are mechanically stable at nonzero overcompression[32,35] ($\Delta\varphi>0$) for all aspect ratios. Interestingly, the density for disordered jammed packings of self-dual ellipsoids seems to surpass that of ellipsoids with other skewness $\beta$[32,36,37].

In this work, we discover via Monte Carlo simulations a novel crystalline packing of self-dual ellipsoids ($\beta=1/2$) that is denser than the corresponding SM2 packing for aspect ratios $\alpha$ in (1.365, 1.5625), attaining a maximal $\varphi \approx 0.75806\ldots$ at $\alpha = 93/64$. The smallest periodic unit cell contains 24 ellipsoids with a quasi-square-triangle tiling arrangement. Henceforth, we refer to this new packing as the square-triangle (SQ-TR) crystal. The SQ-TR crystal of other biaxial ellipsoids with the skewness $\beta$ deviating from 1/2 is readily constructed. The degree of symmetry (order) of the SQ-TR packing varies with the aspect ratio $\alpha$. As a new solid phase, its thermodynamic stability is studied via both compression/decompression Monte Carlo simulations as well as free energy calculations. We show that the SQ-TR phase is thermodynamically stable at high densities over the aforementioned $\alpha$ range and possesses a lower kinetic barrier for crystallization compared to the SM2 phase. A modified phase diagram of hard ellipsoid is proposed.

**Results**

**The SQ-TR crystal.** The SQ-TR packings of congruent self-dual ellipsoids with $\beta = 1/2$ are obtained by compressing a periodic random packing with a low initial density ($\varphi = 0.1$) via the adaptive shrinking cell (ASC) algorithm[13], followed by compression-relaxation process[14] to achieve the maximum density (see Methods for details). Figure 1a shows the density $\varphi$ of the SM2 crystal[5] and the SQ-TR crystal as a function of the aspect ratio $\alpha$. It can be seen that the SQ-TR crystal is denser than the SM2 crystal for $\alpha$ in (1.365, 1.5625). This is a significant breakthrough in the pursuit of optimal packing of ellipsoids since the discovery of SM2 crystal by Donev *et al.*[5] in 2004. Unlike the SM2 crystal with a relatively simple 2-particle basis, the SQ-TR crystal possesses a more complex unit cell and its packing density depends on both the aspect ratio $\alpha$ and the skewness $\beta$ (Supplementary Fig. 1). For a given $\alpha$, the packing density is a non-monotonic function of $\beta$ and the maximum achievable density might correspond to a value of $\beta$ deviating from 1/2.

The inset in Fig. 1a shows a motif of the SQ-TR crystal of self-dual ellipsoids with $\alpha = 93/64$,

where the maximum density $\phi = 0.75806$ is obtained. The maximum density of the SM2 crystal at this aspect ratio is about 0.75085. The SQ-TR crystal possesses a unit cell with 24 ellipsoids, which are packed in two sub-layers. The unit cell of the densest SQ-TR crystal associated with $\alpha = 93/64$ is a cuboid defined by the lattice vectors $\mathbf{a}_1 = (L_x, 0, 0)$, $\mathbf{a}_2 = (0, L_y, 0)$ and $\mathbf{a}_3 = (0, 0, L_z)$ ($L_z \approx 0.6L_x \approx 0.6L_y$, detailed data are available in Supplementary Data). By projecting this packing structure to the XY plane, as shown in Fig. 1b, the projected centroids of the four red ellipsoids in the upper layer form a quasi-square-triangle tiling in 2D. Note that the projected centroids of the four red ellipsoids in the lower layer coincide with those in the top layer. We therefore refer to this configuration as the square-triangle (SQ-TR) crystal.

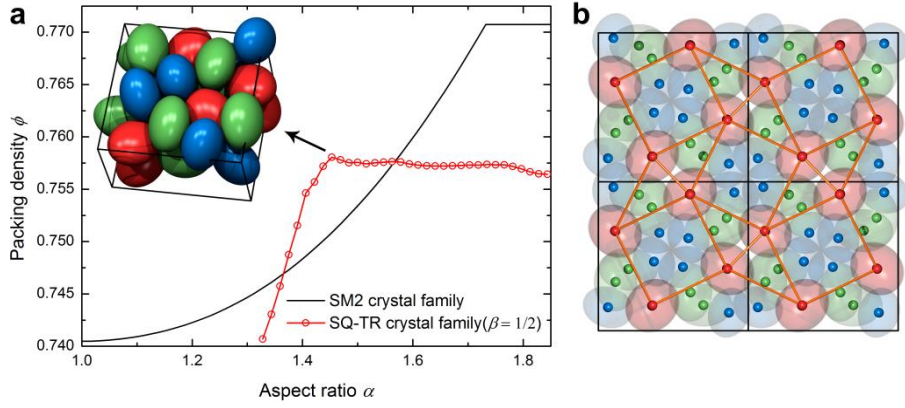

**Figure 1 | Density and visualization of the new dense packing structure of self-dual ellipsoids.** (**a**) The density $\phi$ of the SM2 crystal family[5] and the SQ-TR crystal family as a function of the aspect ratio $\alpha$ ($\beta = 1/2$). The inset is a motif of the densest point of the SQ-TR crystal at $\alpha = 93/64$. (**b**) The projection of the SQ-TR crystal with 2×2×1 unit cells along the Z axis. The ellipsoids are made subtransparent and the centroids are enlarged for better visualization. The square-triangle tiling structure is highlighted in brown.

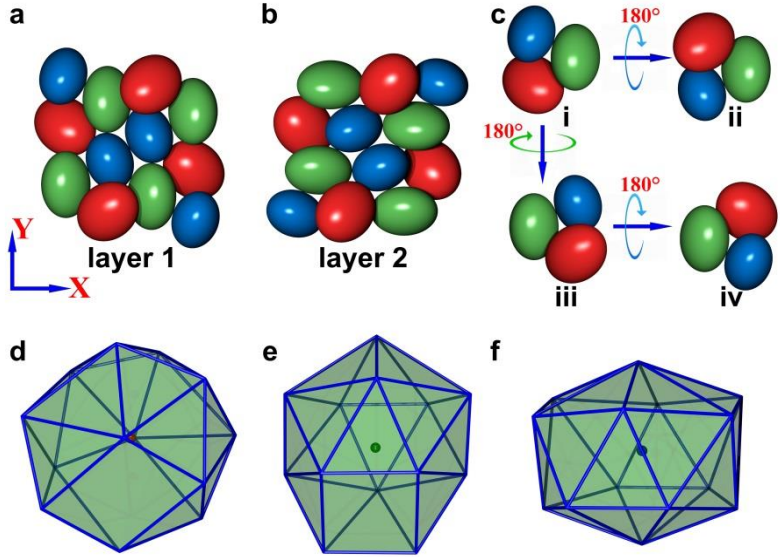

**Figure 2 | Structural characterization of the SQ-TR crystal at $\alpha = 93/64$.** The crystal structure of the inset in Fig. 1a can be partitioned as two layers (**a**) and (**b**), which are equivalent despite the translation and rotation. The layer 1 (**a**) is subdivided into four replica clusters (**c i, ii, iii, iv**). Corresponding ellipsoids in different replicas are

rendered with the same color. (**d**) Coordinates identified around the red ellipsoid in (**c**). Vertexes of the polyhedron correspond to the centroids of coordinated ellipsoids. (**e**) Coordinates identified around the green ellipsoid in (**c**). (**d**) Coordinates identified around the blue ellipsoid in (**c**).

In spite of the relatively complex unit cell structure, a high degree of symmetry is observed for the densest SQ-TR crystal. The 24-particle basis shown in Fig. 1a can be divided into two sub-layers: upper (Fig. 2a layer 1) and lower (Fig. 2b layer 2). Layer 2 can be obtained by first rotating layer 1 by 90 degrees counterclockwise around the Z axis and then taking the mirror imaging the layer about the XY plane. Furthermore, each layer can be subdivided into four three-particle cluster (Fig. 2c) bases, which are symmetric images of one another. In particular, the clusters shown in Fig. 2c (ii), (iii), and (iv) are images of the cluster basis shown in Fig. 2c (i) obtained by rotating the original cluster by 180 degrees around the X, Y, Z axes, respectively. These four clusters are then brought to appropriate contact by translations forming layer 1 in Fig. 2a. We identify the local contact configurations of the particles (Supplementary Note 1) in the SQ-TR crystal, and find only three types of polyhedra formed by the centroids of the contacting neighbors of a central ellipsoid (Fig. 2d-f). Both the polyhedra (Fig. 2d, e) centered at the red and green ellipsoids in the three-particle cluster possess 13 vertexes, while the coordination number $N_c$ of the remaining blue particle in the basis is 14 (Fig. f). Additionally, the polyhedron of Fig. 2d has 18 triangular faces and 2 quadrangular faces, the polyhedron of Fig. 2e has 20 triangular faces and one quadrangular face, and all the faces of the polyhedron of Fig. 2f are triangular.

We note that the degree of symmetry of the SQ-TR crystals varies with the aspect ratio $\alpha$, and is also sensitive to the skewness $\beta$ (Supplementary Fig. 2). The packing structures associated with other $\alpha$ and $\beta$ values possess lower symmetry than this densest SQ-TR crystal ($\alpha = 93/64$, $\beta = 1/2$).

**Thermodynamic stability of the SQ-TR phase.** To understand the thermodynamic stability of the newly discovered SQ-TR phase, especially with respect to its competitor SM2 phase at high densities, we further compute its Helmholtz free energy $A(N, V, T)/k_B T$ as function of packing density for different aspect ratios. Specifically, we employ the standard *NVT* Einstein crystal method[18,27,40], by constructing a reversible path between the actual hard-particle crystal and an ideal Einstein crystal, whose free energy is known analytically. The free energy difference between these two systems is calculated incrementally via MC simulations, which then allows us to obtain the absolute value of the free energy for the hard-particle crystal via a path integral. In our simulations, N = 384 particles are used and the details of this method are given in the Method section. Our free energy calculations indicate that for $\alpha$ in (1.365, 1.5625), for which the SQ-TR crystal is denser than the SM2 crystal, the SQ-TR phase is also thermodynamically stable at high densities (see Supplementary Fig. 3).

To further elucidate the entire phase diagram of the self-dual ellipsoids incorporating the newly discovered SQ-TR phase, we carry out decompression Monte Carlo simulations[40], respectively starting from the high density SQ-TR and SM2 crystal families. The cubatic order parameter $\bar{P}_4$ (ref. 41) and nematic order parameter $S$ (ref. 42, Supplementary Note 2) are employed to evaluate the orientational orders in the system, and the translational order is measured using the global and local bond-order orientational parameters $Q_6$ (ref. 43) and $q_6$ (ref. 44). Figure 3b illustrates the reduced pressure $p^{*[40,45]}$ as a function of packing density $\phi$, in which the approximate boundaries of the coexistence regions are marked. Significant jump of the translational order $q_6$ occurs between the isotropic fluid and solid phases (Fig. 3a) for both aspect ratios (The aspect ratios $\alpha_1 = 7/5$, $\alpha_2 = 93/64$ are chosen as examples for demonstration).

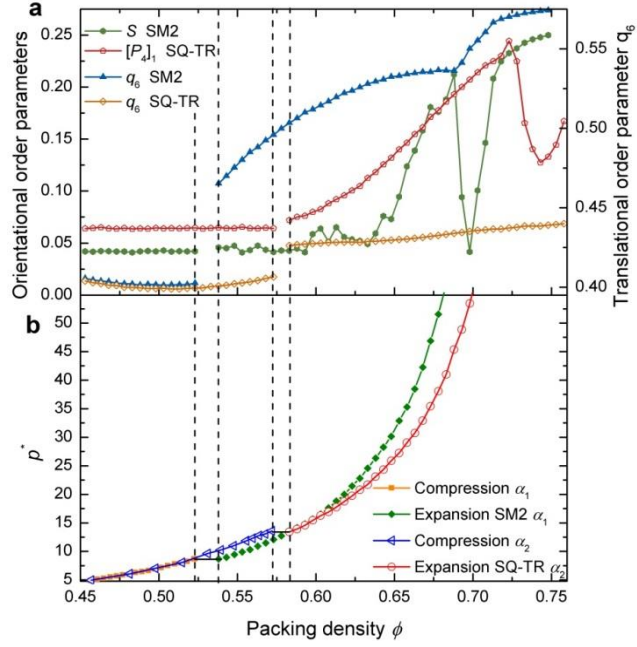

**Figure 3 | Phase and structural behavior of SM2 crystal at $\alpha_1 = 7/5$ and SQ-TR crystal at $\alpha_2 = 93/64$.** (**a**) Orientational order parameter $S$ and translational order parameter $q_6$ of the expansion-run from the SM2 crystal with aspect ratio $\alpha_1 = 7/5$, and Order parameters $[P_4]_1$ and $q_6$ of the expansion-run from the SQ-TR crystal with $\alpha_2 = 93/64$. (**b**) Equations of state with marked phase boundaries (dashed vertical lines) of the coexistence regions using the expansion-runs from the SM2 and SQ-TR crystals, and the compression-runs from isotropic phases. $p^* = pv/(k_B T)$, where $p$ is the pressure, $v$ is the volume of a particle, $k_B$ is the Boltzmann's constant, and $T$ is the temperature.

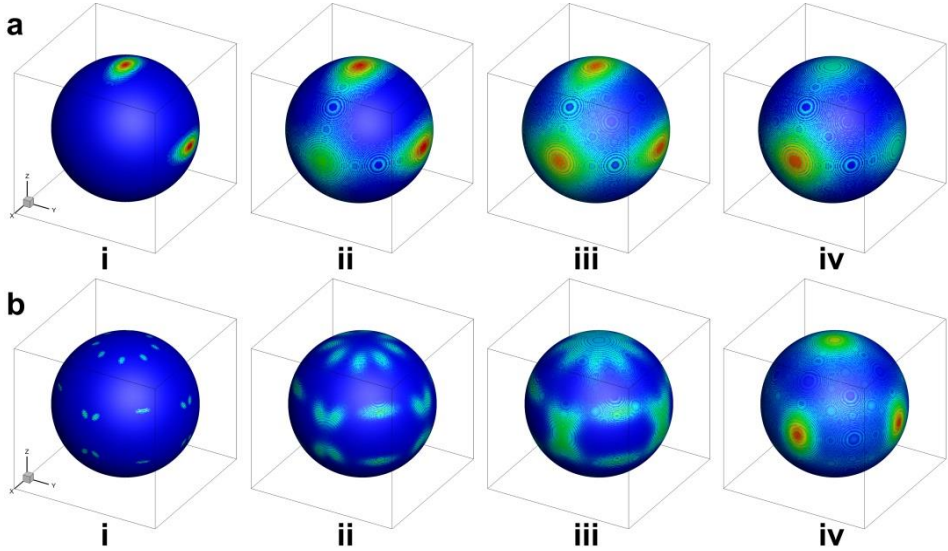

**Figure 4 | Orientation distribution of particles in the systems.** (**a**) The orientation distribution of ellipsoids ($\alpha = 7/5$) during the decompression process from the dense SM2 crystal with density $\phi = 0.718$ (**i**), $\phi = 0.698$ (**ii**), $\phi = 0.693$ (**ii**), $\phi = 0.688$ (**iv**). (**b**) The orientation distribution of ellipsoids ($\alpha = 93/64$) during the

decompression process from the dense SQ-TR crystal with density $\phi = 0.753$ (**i**), $\phi = 0.743$ (**ii**), $\phi = 0.733$ (**iii**), $\phi = 0.723$ (**iv**).

Interestingly, the orientational order parameter $S$ exhibits a non-monotonic behavior, with an initial decrease upon decompression from the dense SM2 crystal, followed by an increase as the decompression proceeds, and finally decreasing again to the minimal value associated with a randomly oriented system. Similar trend of $[P_4]_1$ is observed during the decompression of the SQ-TR crystal. For better understanding the observed non-monotonic behavior of the order metrics, we calculate the distribution of ellipsoid orientation during the expansion-run from the SM2 crystal, as visualized in Fig. 4a. The SM2 phase transits from the two-orientation (along the Y and Z axis) structure to the smectic phase (aligning along the X axis mainly but also with translational order), going through a very narrow region of cubatic phase. The smectic phase here is referred to as the stretched FCC (sFCC) phase. In contrast, the SQ-TR structure evolves from the 24-orientation dominated structure to the cubatic phase with centers of ellipsoids preserving the square-triangle tiling (Fig. 4b).

In addition, our decompression simulations of small system (N = 24) from a low-density state ($\phi = 0.1$) show that there is a large jump of $Q_6$ from around 0.15 to 0.57 during the compression process of the SM2 crystallization branch, compared to a relatively small decrease of $Q_6$ in the densification process of the SQ-TR crystal (see Supplementary Text). This indicates that the crystallization progress of the SQ-TR solid only involves a local structural arrangement compared to the global arrangement for the formation of SM2 solid crystal, thus, the crystallization of SQ-TR structure possesses a lower kinetic barrier.

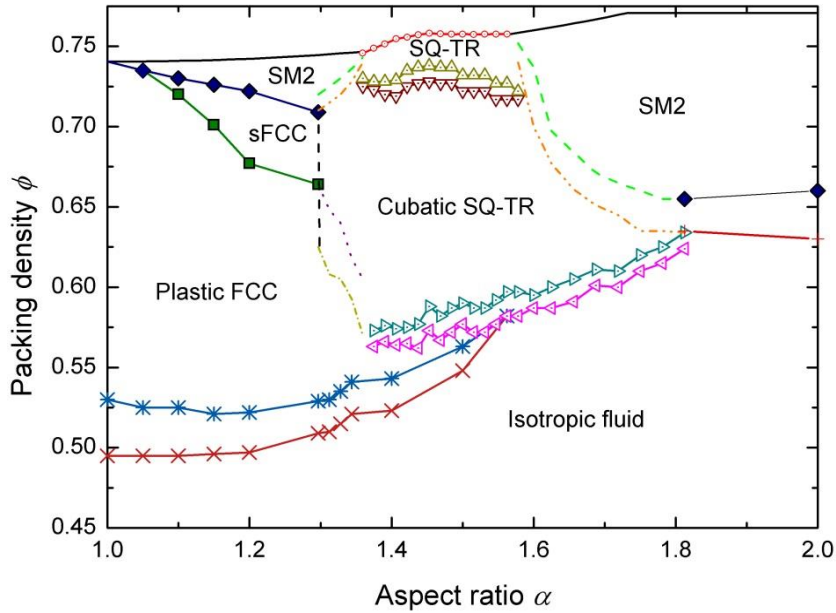

**Figure 5 | Phase diagram of self-dual ellipsoids in the $\phi$ versus $\alpha$.** The dark solid line is the maximally achievable density of the SM2 arrangement, and the red solid line with circles is the maximum density of the SQ-TR crystal. There are several main transitions associated with decompression from SM2 crystal: isotropic-plastic FCC fluid-solid, plastic FCC-stretched FCC solid-solid, sFCC-SM2 solid-solid, and isotropic-SM2 fluid-solid. Three major phases are observed associated decompression of the SQ-TR crystal: isotropic fluid, cubatic SQ-TR solid, and SQ-TR solid. The dashed lines obtained from free energy calculations

mark the boundaries of the coexistence region.

Our free energy calculations and MC simulations enable us to construct the most comprehensive and accurate phase diagram of self-dual ellipsoids to date, which is shown in Fig. 5. The phase regions deriving from the SM2 crystals are similar to that of ellipsoids of revolution in ref. 28 despite the divergence of the corresponding $\phi$-$\alpha$ boundaries for each phase. The intermediate cubatic SQ-TR phase with the preservation of translational order arises from the rearrangement of orientations of particles in the initial SQ-TR crystal and is stable over a wide density range.

**Discussion**

We reported a novel crystalline packing of self-dual ellipsoids ($\beta=1/2$), which are widely used models for the simulation of biaxial nematics[29-31]. Our new crystal is significantly denser than the corresponding SM2 packing for aspect ratios $\alpha$ in (1.365, 1.5625). The smallest periodic unit cell possesses a quasi-square-triangle tiling arrangement and thus, the new packing was referred to as the square-triangle (SQ-TR) crystal. The SQ-TR crystal of other biaxial ellipsoids with the skewness $\beta$ deviating from 1/2 is readily constructed. The degree of symmetry (order) of the SQ-TR packing varies with the aspect ratio $\alpha$ and the skewness $\beta$.

The fundamental cell of the SQ-TR crystal contains 24 particles, which is significantly more complex than that of the SM2 crystal and the densest known packings of other nonspherical shapes. This result is not unexpected as the particles themselves, i.e., self-dual ellipsoids, possess lower symmetry than the ellipsoids of revolution, and suggests an organizing principle that particles with lower symmetry requires more complex arrangements in their fundamental cell to achieve optimal packing. Interestingly, the 24-basis fundamental cell for the most symmetric SQ-TR crystal associated with $\alpha = 93/64$ can be constructed by applying proper symmetry operations (e.g., 180 and 90 degree rotations) to three-particle basis, in contrary to the two-particle basis in the case of SM2 crystal.

As a new solid phase, thermodynamic stability of SQ-TR phase has been studied via both compression/decompression Monte Carlo simulations as well as free energy calculations. We show that SQ-TR phase is thermodynamically stable at high densities for $\alpha$ in (1.365, 1.5625). We also constructed the most comprehensive and accurate phase diagram of self-dual ellipsoid known to date. We note that our MC compression simulations suggest that the SQ-TR phase possesses a lower kinetic barrier for crystallization compared to the SM2 phase, and the crystallization process involves only local particle re-arrangements. The rich structures and low kinetic barrier associated with the SQ-TR phases open fruitful new avenues for engineering nanoparticles and colloids with similar shapes for targeted self-assembly and other applications.

**Methods**

**ASC algorithm.** We use the ASC optimization scheme in ref. 13 to obtain dense packings of ellipsoids. Simulations are initialized at a low packing density ($\phi = 0.1$) in a random configuration with periodic fundamental cell. The unit cell is preset to be a cuboid, whose three edge lengths $L_x$, $L_y$, $L_z$ satisfy that $L_x = L_y = 2L_z$. Trial moves with a Metropolis acceptance rule are carried out before each compression. Each trial move is equally likely to be a random translation or a random rotation of the particle. The translation and rotation magnitudes are adjusted to maintain the rate of successful trial moves around 0.3. The deformation/compression/expansion of the boundary corresponds to collective motions of the particle centroids. The Perram-Wertheim (PW) overlap potential[11,46] is used here to detect overlaps between ellilsoids. Slow compression is applied to particulate systems to obtain dense structures. Note that the compression rate may vary with the aspect ratio $α$ and the skewness $β$.

**Compression-relaxation process.** To obtain the maximum density for a given shape of ellipsoids, dense packings obtained by the ASC algorithm are further optimized through a modified compression progress, as described in ref. 14. For the modified compression scheme, cell changes are always accepted even if they result in new overlaps and a different criterion is used to ensure that the amount of overlaps remain small. If the rate of successful trial moves is larger than 0.3, a compression move is applied, otherwise an expansion move is applied. At the end, a few hundred cycles of variable-shape *NVT* Monte Carlo[47] are carried out to remove all minor overlaps generated by this method.

**Free energy calculation of a solid phase.** We employ the standard *NVT* Einstein crystal method[18,27,40] to calculate the Helmholtz free energy. The free energy difference between the actual hard-particle crystal and an ideal Einstein crystal is calculated incrementally via MC simulations, which then allows us to obtain the absolute value of the free energy for the hard-particle crystal via a path integral, i.e.,

$$\frac{A(N,V,T)}{k_B T} = \frac{A_E(N,V,T)}{k_B T} - \int_0^{\lambda_{max}} \left\langle \frac{\partial U_E(\lambda)}{\partial \lambda} \right\rangle, \tag{1}$$

where $V$ is the volume of the system, and $\lambda_{max}$ is the maximum coupling constant that is sufficiently strong to suppress particle collisions ($\lambda_{max} = 8000$ in this work). $A_E$ is the free energy of the ideal Einstein crystal given by

$$\frac{A_E(N,V,T)}{k_B T} = -\frac{3(N-1)}{2} \ln\left(\frac{\pi k_B T}{\lambda_{max}}\right) + N \ln\left(\frac{\Lambda_t^3 \Lambda_r}{\sigma^4}\right) + \ln\left(\frac{\sigma^3}{VN^{1/2}}\right)$$
$$- \sum_{i=1}^N \ln\left\{\frac{1}{8\pi^2} \int d\theta d\phi d\chi \sin\theta \exp\left[-\frac{\lambda_{max}}{k_B T}\left(\sin^2\psi_{ia} + \sin^2\psi_{ib}\right)\right]\right\}, \tag{2}$$

and the coupling potential $U_E$ is

$$U_E(\lambda) = \lambda \sum_{i=1}^N \left[(\mathbf{r}_i - \mathbf{r}_{i0})^2 / \sigma^2 + \left(\sin^2\psi_{ia} + \sin^2\psi_{ib}\right)\right], \tag{3}$$

where $\mathbf{r}_i$-$\mathbf{r}_{i0}$ is the displacement of particle $i$ with respect to its reference lattice site, and $\psi_{ia}$, $\psi_{ib}$ are the angles between the two axes of a particle and the reference orientations of the *i*th particle in the crystal, respectively. Here $\Lambda_t$ and $\Lambda_r$ are the translational and rotational thermal wavelengths of the particles, respectively, and both are set to 1. $\sigma = 1/v^{1/3}$ is the unit length of the particle with volume $v$.

**Acknowledgements**

We thank Duyu Chen for his help with free energy calculation and Corey S. O'Hern for his suggestion in inducing particle rearrangement. This work was supported by the National Natural Science Foundation of China (Grant Nos.11272010 and 11572004).

**Supplementary Figures**

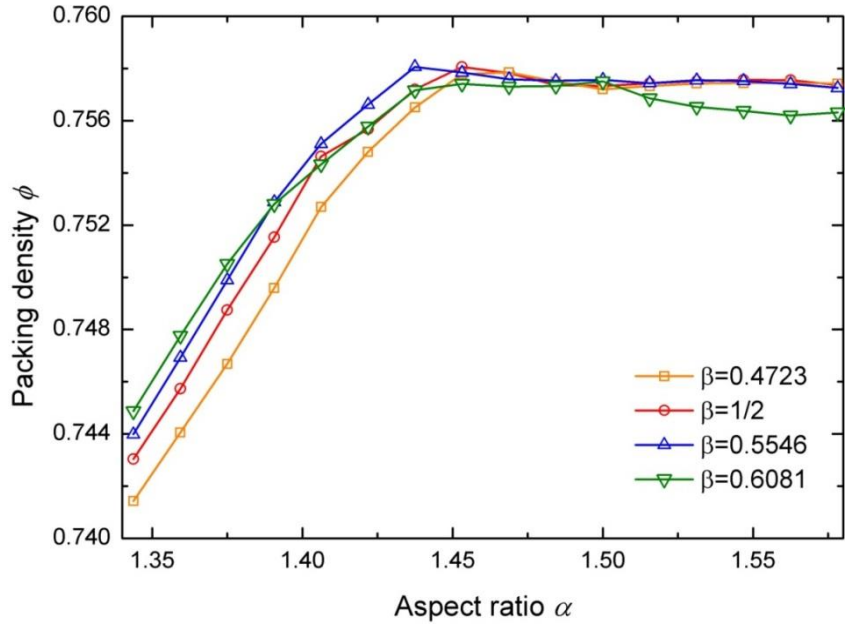

**Supplementary Figure 1. Packing density of the SQ-TR crystal.** The density $\phi$ of the SQ-TR crystal family as a function of the aspect ratio ratio $\alpha$ with various skewness $\beta$. For a given aspect ratio, the density varies with the change of $\beta$ and the best skewness for ellipsoids packing densest seems to be larger than 0.5 when $\alpha < 93/64$.

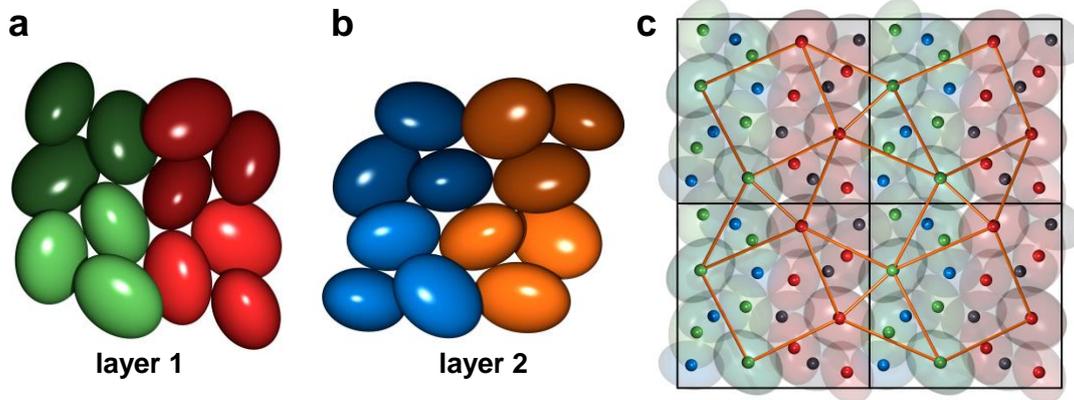

**Supplementary Figure 2. Visualization of the SQ-TR structure of self-dual ellipsoids at $\alpha = 25/16$.** For the structure with lower degree of order compared to the densest point, the configuration can be divided into two different layers: layer 1 (**a**) and layer 2 (**b**). The whole arrangement is constructed by four different replica pairs. Each replica pair is rendered by the same color with different brightness scales. In each pair, the lighter colored cluster can be obtained by rotating the darker colored one by 180 degrees around the Y axis. (**c**) The projected pattern along the cell vector $\mathbf{a}_2$. The structure presents a quasi square-triangle tiling not as symmetric as the SQ-TR tiling motif at the densest packing point.

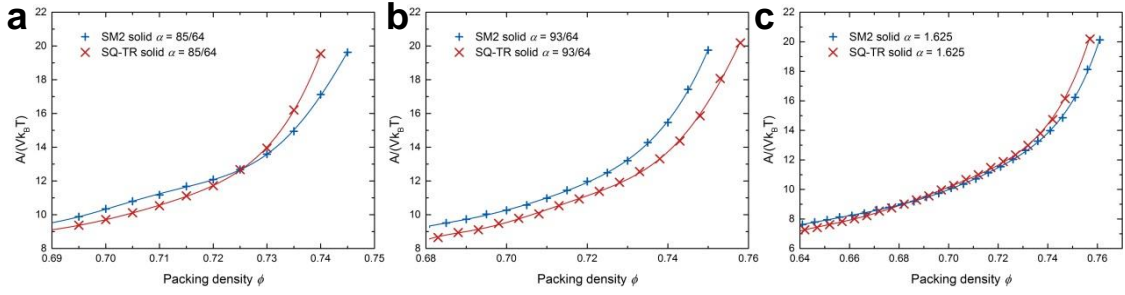

**Supplementary Figure 3. Reduced Helmholtz free energy per unit volume $A/(Vk_BT)$ as a function of packing density.** Three aspect ratios are chosen as examples: (**a**) $\alpha = 85/64$, (**b**) $\alpha = 93/64$, and (**c**) $\alpha = 1.625$. For $\alpha$ within (1.365, 1.5625), in which the SQ-TR crystal is denser than the SM2 crystal, the SQ-TR phase has lower free energy at high densities. For $\alpha$ outside this range but still close to the bounds, the SM2 phase is more stable at high density region, while the free energy of the SQ-TR phase at low densities become smaller. The boundaries of the coexistence region obtained from free energy calculations are marked with dashed lines in Fig. 5.

## Supplementary Notes
### Supplementary Note 1
A cutoff distance for searching coordinates around a particle is defined by the position of the first minimum in the radial distribution.

### Supplementary Note 2
The direction of the longest axis of an ellipsoid is chosen as the orientation of the particle for calculating the nematic order parameter $S$.

## Supplementary Text
**Crystallization progress.** The dense packings are obtained by compressing random initial systems with a relatively low density ($\varphi = 0.1$). Two dense crystalline branches ($N = 24$), crystallizing into the SM2 crystal or the SQ-TR crystal, are observed during the compression. Packings of self-dual ellipsoids with $\alpha = 25/16$, as shown in Supplementary Fig. 4a, are chosen as examples for demonstration, for maximum densities of the SM2 and SQ-TR crystals at this aspect ratio are close (Fig. 1a). It can be seen that there is a large jump of $Q_6$ from around 0.15 to 0.57 during the compression process of the SM2 crystallization branch, compared to a relatively small decrease of $Q_6$ in the densification process of the SQ-TR crystal. We therefore conjecture that the crystallization progress of the SQ-TR solid is a local optimization evolution in contrary to the global optimization of the SM2 solid forming. The locally and globally favorable structures are incompatible, which makes the intermediate state pack looser than the two types of crystals (Supplementary Fig. 4a). The coordination number $N_c$ of particles in the SM2 crystal is 12, while 2/3 of particles in the SQ-TR system are coordinated by 13 and the rest are coordinated by 14. The intermediate state contains all three values of $N_c$ and clusters formed by coordinates are variants of that formed in the two crystals.

In addition, the dense SQ-TR crystal is obtained more frequently in our Monte Carlo simulations implying that the SQ-TR crystal wins the competition with the globally optimized SM2 structure for the given particle number $N = 24$ with the preset cell condition (Methods) .

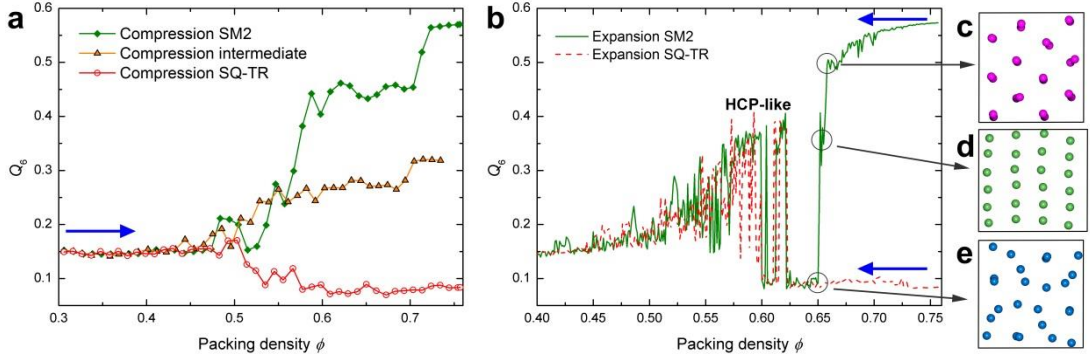

**Supplementary Figure 4: Competition between the SM2 structure and the SQ-TR structure ($N = 24$) at $\alpha = 25/16$.** (**a**) Crystallization progress of the SM2 solid and the SQ-TR solid from random initial configurations. The maximum densities of the two crystals at this aspect ratio are comparable, while the $Q_6$ value of the final two crystal states is about 0.57 and 0.08, respectively. The maximally achievable density of the intermediate state is smaller than that of both crystals. (**b**) Expansion-runs from the dense SM2 and SQ-TR crystals. The branch of the SM2 decompression process undergoes a transition from the SM2 structure (**c**) to the HCP-like structure (**d**) and then to the SQ-TR structure (**e**). Motifs in (**c-e**) are projected patterns of centroids of ellipsoids along the cell vector $\mathbf{a}_2$.

**Direct observation of solid-solid phase transition.** In the phase diagram of self-dual ellipsoids in Fig.5, no spontaneous SM2-SQ-TR solid-solid transition is observed in large systems. Here in small systems ($N = 24$), we decompress the dense SM2 and SQ-TR crystals (Supplementary Fig. 4b) by $\Delta\varphi = 0.001$ per expansion, and observe solid-solid phase transitions, respectively. In the expansion-run from the SM2 crystal, the solid phase first transits from the SM2 state (Supplementary Fig. 4c) to a derived medium state of quasi hexagonal close packed (HCP-like) structure (Supplementary Fig. 4d), and then quickly to the SQ-TR state (Supplementary Fig. 4d), when the system is diluted to a certain extent. Afterwards, jumps between the SQ-TR solid and the HCP-like solid are detected as the decompression proceeds. In similar manner, these jumps of the SQ-TR-HCP-like solid-solid transition arise during the decompression of the SQ-TR solid (dashed line in Supplementary Fig. 4b). The presence of these solid-solid transitions may be attributed to the break of constraints and the deformation of cell. During the expansion, the free volume per particle enlarges and particles have more freedom of motion. When the free volume increases to a certain extent, the interparticle constraints become week enough for particle trial moves to break. Moreover, the deformation of the adaptive fundamental cell (collective motions of particles) promotes the break progress, and makes particles rearrange into a new stable solid.

Notably, for large systems, particle trial moves are less efficient to break interparticle constraints and the effect of the cell deformation becomes smaller, which makes the solid-solid transition between the SM2 crystal and the SQ-TR crystal invisible. However, the solid-solid transition may appear at low density if an inducing factor is put inside the system. As shown in Supplementary Fig. 5a, we replace 24 particles of a plastic FCC crystal (N = 384, $\varphi = 0.6$, $\alpha = 93/64$) with a SQ-TR cluster (N =24), which is fixed as an inducing factor. The system then evolves into an approximant of the SQ-TR crystal after a long equilibration progress (Supplementary Fig. 5). This particle rearrangement behavior is not observed in high density region.

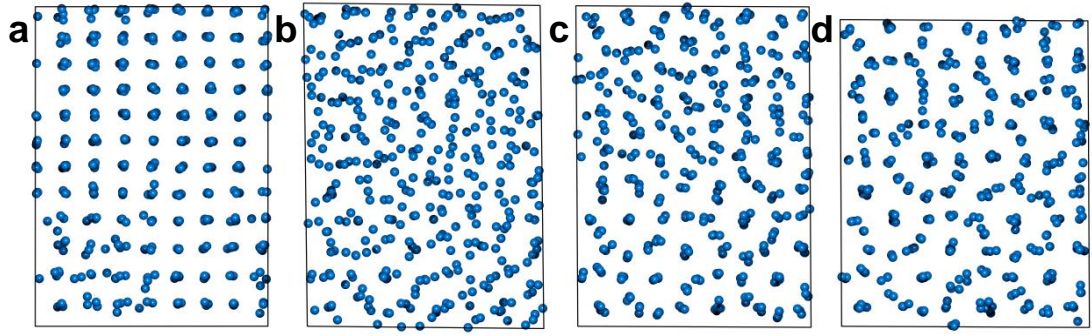

**Supplementary Figure 5. Evolution of particle rearrangement with a SQ-TR cluster preset.** Particles in the system rearranges from the initial plastic FCC crystal with 24 ellipsoids fixed as a SQ-TR cluster (**a**) to an approximant of the SQ-TR crystal (**d**). Motifs (**b**) and (**c**) are intermediate states. Only centroids of particles are shown for better visualization.

**Supplementary Data**

In the following, we list the detailed data of the densest packing point of self-dual ellipsoids at $\alpha = 93/64$. In the first line, the number of particles per unit cell is given. The second line presents the lengths of the three semiaxes. Three cell vectors $\mathbf{a}_1 = (a_{11}, a_{12}, a_{13})$, $\mathbf{a}_2 = (a_{21}, a_{22}, a_{23})$ and $\mathbf{a}_3 = (a_{31}, a_{32}, a_{33})$ are listed in the next three lines, and following are the coordinates of centroids and three unit vectors along the semiaxes belonging to each ellipsoid in the structure.

```
24
1.2054563451241194    1    0.8295613557843402
6.0612425446099198    0    0
0    6.0612389571219083    0
0    0    3.60971734418704140.6400273340148365    5.53956715718213481.6242199477973345
0.9595214569698769    0.0610861313021496    -0.2749310055949362
-0.2699234559414653    0.4780509094345219    -0.8358281256355010
0.08037351062780710.8762053480618247    0.4751886847975574
5.7554471785578736    1.9890828459742229    3.4945729351564285
-0.0610832475464592    0.9595215536886753    0.2749313088164866
0.4780513242982217    0.2699223196166387    -0.8358282552469567
-0.8762053227024391    0.0803761720491898    -0.4751882813010160
1.2484791872543117    2.8160394994951830    3.4520167713726804
-0.1495158545993143    0.3616297039596012    0.9202548376172699
-0.2690053346108559    0.8807318926251998    -0.3898043905039277
-0.9514626311722422    -0.3058353970932521    -0.0344030620798510
5.7307850708107102    4.0653051319767615    2.7010628587733483
-0.6789512395672668    0.6140427434445004    -0.4024633192301772
0.6971445702358353    0.7111169080564511    -0.0911163604986683
0.2302491312038197    -0.3424386836258385    -0.9108902706872191
```

| | | |
|---|---|---|
| 2.7248258726894381 | 1.5631242531572440 | 3.4290787038122614 |
| -0.0610834864535167 | -0.9595214763604594 | -0.2749315256249689 |
| -0.4780517418062812 | 0.2699225830794400 | -0.8358279314538072 |
| 0.8762050783385288 | 0.0803762105740451 | -0.4751887255216833 |
| 0.2140706914841905 | 2.5089477003628931 | 1.6897143375138521 |
| 0.9595214130681327 | -0.0610861157986623 | -0.2749311622272463 |
| -0.2699235517379630 | -0.4780512725961911 | -0.8358278869890391 |
| -0.0803737128867213 | 0.8762051509972987 | -0.4751890139387678 |
| 2.4177338790356706 | 2.5975327696764827 | 1.6667758482631805 |
| -0.3616320335306312 | -0.1495143412642078 | -0.9202541679833756 |
| 0.8807308701215609 | 0.2690060867113748 | -0.3898061816496182 |
| 0.3058355869790988 | -0.9514626562537535 | 0.0344006777077259 |
| 5.7468729959096407 | 4.0492173762555916 | 0.8962041873056167 |
| 0.6140501760868782 | -0.6789448442499513 | -0.4024627679537223 |
| -0.7111105508620399 | -0.6971514239375036 | 0.0911135360059493 |
| -0.3424385572733881 | 0.2302472377926815 | -0.9108907968275667 |
| 1.1684675169798451 | 1.0185979265114791 | 2.4177300921227887 |
| -0.6140500466500207 | -0.6789448778071319 | 0.4024629088113130 |
| -0.7111105345086910 | 0.6971514076164942 | 0.0911137883636512 |
| -0.3424388232983396 | -0.2302471882095752 | -0.9108907093284033 |
| 1.1845555104120651 | 1.0346856242427713 | 0.6128714190648815 |
| 0.6789511661472983 | 0.6140429218024645 | 0.4024631710416187 |
| -0.6971446729365904 | 0.7111168204782639 | 0.0911162586140058 |
| -0.2302490368876625 | -0.3424385457857904 | 0.9108903464196837 |
| 4.1990888652745850 | 2.5336092014380438 | 0.8962041911367922 |
| -0.6140501328313107 | 0.6789448410680432 | -0.4024628391939850 |
| 0.7111106274931179 | 0.6971513664337132 | 0.0911133768702659 |
| 0.3424384754612338 | -0.2302474209520004 | -0.9108907811272629 |
| 5.4483551564927941 | 0.9546742836037321 | 1.6471584587049386 |
| -0.3616321654490824 | 0.1495144268888981 | 0.9202541022243170 |
| -0.8807308383128911 | 0.2690059755961630 | -0.3898063302243633 |
| -0.3058355226263838 | -0.9514626742176964 | 0.0344007530601472 |
| 4.1905146681824155 | 4.5937437232507552 | 3.4945728820862687 |
| 0.0610833862098201 | -0.9595215186230115 | 0.2749314001348235 |
| -0.4780513025859014 | -0.2699224667261532 | -0.8358282201595527 |
| 0.8762053248203294 | -0.0803760959791456 | -0.4751882901186738 |
| 4.4976066532881234 | 5.6281522765497014 | 1.6471584164000148 |
| -0.3616322229321766 | 0.1495143922070840 | -0.9202540853355294 |
| -0.8807307515294711 | 0.2690061846339154 | 0.3898063821452577 |
| 0.3058357046947748 | 0.9514626206639659 | 0.0344006184813508 |
| 4.2791004718577330 | 0.7361676477719215 | 3.4716348640465733 |
| -0.1495159437783265 | -0.3616298648918385 | -0.9202547598601570 |
| 0.2690052754454870 | 0.8807318313420951 | -0.3898045698854334 |
| 0.9514626338948066 | -0.3058353833117272 | -0.0344031095641839 |
| 4.2151767769622976 | 2.5175214385260873 | 2.7010628593909494 |
| -0.6789512580340726 | 0.6140427687380051 | 0.4024632496234353 |
| 0.6971446020828560 | 0.7111168601885994 | 0.0911164904181796 |
| -0.2302489803594489 | 0.3424387378087755 | -0.9108902885080010 |
| 1.1598934133472363 | 5.0197023388470496 | 3.4290787137696443 |
| -0.0610833815743341 | -0.9595214794036656 | 0.2749315382469842 |
| -0.4780516739568380 | 0.2699225507943907 | 0.8358279806056423 |
| -0.8762051225751558 | -0.0803762826126596 | -0.4751886315921553 |
| 3.6706486248270203 | 4.0738788716448582 | 1.6897142876851443 |
| 0.9595213751178273 | -0.0610862600663123 | 0.2749312627119833 |
| -0.2699237074546349 | -0.4780512722312184 | 0.8358278369002112 |
| 0.0803736433000494 | -0.8762051411506525 | -0.4751890438997625 |
| 5.6668613436063540 | 5.8466589283805073 | 3.4716348458192989 |
| -0.1495159624955905 | -0.3616297574812628 | 0.9202547990577638 |
| 0.2690053300086960 | 0.8807318567361917 | 0.3898044747487853 |
| -0.9514626155130495 | 0.3058354371397727 | -0.0344031390082285 |
| 2.7162517581759560 | 5.5642286647572332 | 2.4177300736010974 |
| 0.6140500187638738 | 0.6789449068930914 | 0.4024629022064779 |
| 0.7111106494069348 | -0.6971513195482295 | 0.0911135654726902 |
| 0.3424386345690094 | 0.2302473691620993 | -0.9108907345021269 |
| 2.6362401362583978 | 3.7667870973615809 | 3.4520167473434378 |
| 0.1495158303908827 | -0.3616297822204039 | 0.9202548107591081 |
| -0.2690054271824792 | 0.8807318329855054 | 0.3898044613701382 |
| -0.9514626087674236 | -0.3058354762168124 | 0.0344029773124054 |
| 2.7001637695767151 | 5.5481409748554462 | 0.6128714028443526 |
| -0.6789512321685134 | -0.6140428348413846 | 0.4024631923975408 |
| -0.6971445683088741 | 0.7111169340865955 | -0.0911161720978984 |
| -0.2302491588455538 | -0.3424384658405413 | -0.9108903456326438 |
| 1.4669854222031407 | 3.9852938138672775 | 1.6667758687641594 |
| -0.3616320979042724 | -0.1495142793211659 | 0.9202541527807177 |
| 0.8807308939714266 | 0.2690059540417740 | 0.3898062194705318 |
| -0.3058354423087261 | 0.9514627035827410 | 0.0344006573791570 |
| 3.2446919317013285 | 1.0432594245317415 | 1.6242199593713009 |
| -0.9595214270859345 | -0.0610861335439700 | -0.2749311094266982 |
| -0.2699235540144786 | 0.4780508377308202 | 0.8358281349529273 |
| 0.0803735381081550 | 0.8762053870142058 | -0.4751886083062946 |